\documentstyle[twocolumn,aps]{revtex}
\begin{document}

%
%

\title{Influence of next-nearest-neighbor electron 
hopping on the static and dynamical properties of the 2D Hubbard model}

\author{ Daniel Duffy and Adriana Moreo}

\address{Department of Physics, National High Magnetic Field Lab and MARTECH,
Florida State University, Tallahassee, FL 32306, USA}

\date{\today}
\maketitle

\begin{abstract}

Comparing experimental data for high temperature 
cuprate superconductors with numerical results for electronic models, it is
becoming apparent that a hopping along the plaquette diagonals
has to be included to
obtain a  quantitative agreement. According to recent estimations the
value of the diagonal hopping $t'$ appears to be material dependent. 
However, the values for $t'$ discussed in the literature
were obtained comparing theoretical results in the 
weak coupling limit
with experimental photoemission data and band structure calculations. 
The goal of
this paper is to study how $t'$ gets renormalized as the interaction
between electrons, $U$, increases. For this purpose, 
the effect of adding a bare 
diagonal hopping $t'$ to the fully interacting two dimensional Hubbard
model Hamiltonian is investigated using numerical techniques. 
Positive and negative values of $t'$ are analyzed.
Spin-spin
correlations, $n(\bf{k})$, $\langle n\rangle$ vs $\mu$, and local magnetic moments
are studied
for values of $U/t$ ranging from 0 to 6, and as a function of the electronic
density.
The influence of the diagonal hopping in the spectral function
$A(\bf{k},\omega)$ is also discussed, and the changes in the gap present in
the density of states at half-filling are studied. We introduce a new
criterion to determine probable locations of Fermi
surfaces at zero temperature 
from $n(\bf{k})$ data obtained at finite temperature. It appears that hole
pockets at ${\bf{k}}=(\pi/2,\pi/2)$ may be induced for negative $t'$ while
a positive $t'$ produces similar features at ${\bf{k}}=(\pi,0)$ and
$(0,\pi)$. Comparisons with the standard 2D Hubbard ($t'=0$) 
model indicate that a
negative $t'$ hopping amplitude appears to be dynamically generated. In
general, we conclude that it is very dangerous to extract a bare
parameter of the Hamiltonian $(t')$ from PES data where renormalized
parameters play the important role.
\end{abstract}

\pacs{74.20.Mn, 74.25.Ha, 74.25.Jb}

\section{ Introduction}
As experimental measurements of the properties of high temperature
superconducting 
materials become more accurate, it has been observed that different
cuprate compounds present slightly different properties. 
Among these qualitative differences are the presence of incommensurate magnetic
correlations, the symmetry of the pairing state, shape of the
Fermi surface (FS), as well as the
behavior of the resistivity with temperature. 

Many models proposed to describe the cuprate
superconductors are particle-hole symmetric and this makes it impossible to
distinguish between electron and hole doped materials. 
The addition of a diagonal hopping term to the one band 
Hubbard model has often been
suggested as a way to handle the different properties between electron
and hole-doped materials since 
such a term breaks particle-hole symmetry. Even
restricted to the family of
hole doped materials the inclusion of a diagonal hopping has also
been proposed in order to reproduce experimental photoemission (PES)
data and  band structure calculations. According to these estimations, 
the diagonal hopping parameter $t'$ seems material
dependent and some of the suggested values are $t'=0.20$ for
$Nd_2CeCuO_4$, $t'=-0.20$ for $La_2SrCuO_4$, and -0.45 for
$YBa_2Cu_3O_7$.\cite{values} 

Recently, the effect of this $t'$ hopping term has
been analyzed in the weak coupling limit for the Hubbard model\cite{Trem,Si}
as well as for the t-J model.\cite{troup,Good} It was reported that the 
Fermi surface 
of the $U-t-t'$ model in the non-interacting case is in agreement with 
experimental measurements 
for appropriate values of $t'$, while the FS of
the $t'=0$ model is not.
The same occurs for the magnetic susceptibility and the sign change of
the Hall coefficient with doping.\cite{Trem} However, it is not clear how a
moderate or strong Coulombic interaction will modify the properties of
weakly interacting electrons. 
In other words, it is not correct to deduce the value of a
$bare$ parameter in a Hamiltonian using data from experiments where the
$renormalized$ parameter plays the key role. It is well known that
strong correlations are fundamental in high $T_c$ models, thus it is
likely that the bare $t'$ is quite different from the renormalized one.

In this paper, we investigate the $U-t-t'$ model in the
intermediate coupling regime using quantum Monte Carlo techniques. We
will compare the results with mean field calculations, as well as with
the non-interacting limit and with experimental results.   
The paper is organized as follows: in Section II the model is introduced
and its symmetry properties are discussed.  Spin-spin
correlation functions and incommensurability are studied in Section III.
Section IV is devoted to the behavior of the density $\langle n \rangle$
as a function of 
chemical potential while local magnetic moments are
discussed in Section V. Angle resolved photoemission
(ARPES) experiments and the shape of
the Fermi surface are studied in Section VI. Dynamical properties 
appear in Section VII and the
conclusions are presented in Section VIII. A mean field analysis is
discussed in the Appendix. 

\section{The model} 

The $U-t-t'$ Hamiltonian is given by
$$
{\rm H=
-t\sum_{<{\bf{ij}}>,\sigma}(c^{\dagger}_{{\bf{i}},\sigma}
c_{{\bf{j}},\sigma}+h.c.)
-t'\sum_{<{\bf{in}}>,\sigma}(c^{\dagger}_{{\bf{i}},\sigma}
c_{{\bf{n}},\sigma}+h.c.)+}
$$

$$
{\rm U\sum_{{\bf{i}}}(n_{{\bf{i}} \uparrow}-1/2)( n_{{\bf{i}}
\downarrow}-1/2)+\mu\sum_{{\bf{i}},\sigma}n_{{\bf{i}}\sigma} },
\eqno(1)
$$
\noindent where ${\rm c^{\dagger}_{{\bf{i}},\sigma} }$ creates an electron at
site ${\rm {\bf i } }$
with spin projection $\sigma$, ${\rm n_{{\bf{i}}\sigma} }$ is the number
operator, the sum
${\rm \langle {\bf{ij}} \rangle }$ 
runs over pairs of  nearest neighbor lattice sites, and the sum
${\rm \langle {\bf{in}} \rangle }$ 
runs over pairs of lattice sites along the plaquette
diagonals. ${\rm U}$ is the
on site Coulombic repulsion, ${t}$ the nearest neighbor hopping
amplitude, ${t'}$ the diagonal hopping amplitude, and $\mu$ is the
chemical potential. 

When ${t'}=0$ the above Hamiltonian is particle-hole symmetric,
i.e., it remains invariant after the particle-hole transformation
operation

$$
{\rm c^{\dagger}_{{\bf{i}},\sigma} }\rightarrow (-1)^{\bf{i}}{\rm c_{{\bf{i}},\sigma}}.
\eqno(2)
$$

\noindent If $t'> 0$ the above transformation maps
the Hamiltonian in its particle-hole symmetric one but with $t'<0$,
and vice versa. This means that the results obtained for negative
(positive) $t'$
below half-filling can be obtained from the results for
positive (negative) $t'$ above half-filling. In fact, this is the reason why
to mimic the effect of electron and hole doping in the t-J model, (which
cannot be doped above half-filling) both signs of
$t'$ are studied below half-filling.\cite{troup,Good} Then, it is
clear that results above and below $\langle n\rangle=1$ with different signs
for $t'$ are redundant and in this paper we will study the 
Hamiltonian (Eq.(1)) only 
below half-filling with positive and negative values for
$t'$. We will set $t=1$ and $|t'|$=0.2 unless stated otherwise.

\section{Magnetic Correlations and Incommensurability}

A spin density wave (SDW) mean field analysis (see Appendix) of the
$U-t-t'$ model predicts the existence of
antiferromagnetism at half-filling for $U>U_c$.\cite{Hirsch} 
At $|t'|=0.2$ it can be shown that $U_c=2.1$.
Thus, within the SDW mean field, adding a $t'$ term appears to decrease 
the strength of
the spin-spin correlations. This result is not unexpected if we
consider the large U limit. In this case a Heisenberg spin-spin
interaction with $J=4t^2/U$ appears between
nearest neighbors sites 
and a $J'=4t'^2/U$ is generated along the diagonals. The
$J'$ interaction introduces frustration and tends to decrease the
tendency to antiferromagnetism. 
In Fig.1-a we present quantum Monte Carlo (QMC) results for 
the spin-spin correlations 
${\rm C({\bf r})=\langle S^z_{\bf i} S^z_{\bf i+r}\rangle (-1)^{|{\bf r}|} }$
as a function of distance, at half-filling, for $U/t=4$, $t'=0$ and
-0.2 on an $8 \times 8$ lattice at $\beta t=6$. 
Clearly, in both cases there is long range order (or, better, a
correlation larger than the system size) but
the strength of the antiferromagnetic correlations decreases when
$|t'|$ increases. 
This is true for either sign of $t'$
since due to the symmetry of the Hamiltonian at $\langle n\rangle=1$ the spin-spin
correlations must be the same for both signs of $t'$. 
A similar effect is observed in Fig.1-b as $U/t$ is increased to 6 at
$\beta t=4$.

Previous quantum Monte Carlo calculations on the 2D Hubbard model have
indicated that
short range incommensurate magnetic correlations develop as the system
is doped away from half-filling.\cite{Hub,tj} After these results were
reported, neutron
scattering experiments revealed short range magnetic incommensuration
in $La_{2-x}Sr_xCuO_4$ (LSCO).\cite{aeppli} The qualitative
agreement between theory and experiment was remarkable. However, at a 
fixed doping, the experimental peaks in the structure factor
S({\bf{k}}) appeared shifted away from ${\bf{k}}=(\pi,\pi)$ by an amount
larger than 
theoretically predicted. Additional confusion arose when experiments
were performed in other materials. It was found that $Nd_{2-x}Ce_xCuO_4$ does
not show incommensurability\cite{Thurston} while for $YBa_2Cu_3O_{6+x}$ (YBCO) 
the result is unclear.\cite{Tran} Several authors suggested that the
addition of a material dependent $t'$ term to the Hubbard Hamiltonian
could explain the quantitative, as well as the qualitative, differences
observed. All previous 
studies of the $U-t-t'$ model have been carried out in the weak
coupling limit ($U/t \leq 4$). There are some disagreements among the 
published results. B\'enard et al.\cite{Trem,Trem2} concentrated
on negative values of $t'$ and used a perturbative approach. They found
that incommensurability starts developing at a finite hole density 
which increases from half-filling as $t'$ becomes more negative. A similar
result, within RPA, was obtained in Ref.\cite{Si} though these authors
expect incommensurability to disappear for $t'=-0.45$ due to a decreasing
intensity of the peaks in the spectral function. Furukawa and
Imada\cite{Imada} studied positive values of $t'$ using a zero
temperature Monte Carlo algorithm for $U/t=4$ and $t'=0.25$. They found
incommensurability, but with the peaks in S({\bf{k}}) 
moving along the diagonals
in the Brillouin zone rather than towards the points $X=(\pi,0)$ and
$Y=(0,\pi)$ as found in the experiment and as it happens for
$t'=0$.\cite{Hub,tj}  
Gooding et al.\cite{Good} performed exact diagonalization studies 
of the  $t-t'-J$ model that should resemble the strong coupling limit of 
the $U-t-t'$ model. To increase the number of available momenta along
particular directions in momentum space they did
not use square clusters. For negative values of $t'$ they observed
incommensurability with S({\bf{k}}) moving along the diagonals, while for
positive values of $t'$ the peak in S({\bf{k}}) remained at $(\pi,\pi)$ for
all the dopings they studied. 

Due to the disparity of all these results and to the
lack of numerical data on large square lattices at more realistic 
couplings larger than
$U/t=4$, we decided to carry out a systematic study of the structure
factor at $U/t=6$ on $8 \times 8$ lattices using quantum Monte Carlo.
We worked at a temperature $T/t=1/4$ since due to sign problems it is not
possible to study lower temperatures away from half-filling. Our results are
presented in Fig.2 where we show S({\bf{k}}) along the
directions $(0,0)-(\pi,0)-(\pi,\pi)-(0,0)$ (or $\Gamma-X-M-\Gamma$)
for $U/t=6$ at different fillings and for positive and negative values
of $t'$. At half-filling, S({\bf{k}}) is independent of the sign of $t'$ due to
the symmetry in the Hamiltonian described in Sec.II.  As the density is 
reduced to $\langle n\rangle=0.9$, the strength of S({\bf{k}}) at the peak is
slightly smaller for 
negative $t'$ (Fig.2-a) than for positive $t'$ (Fig.2-b).
For negative (positive) hopping the maximum in S({\bf{k}}) starts moving
along the lines from $(\pi,\pi)$ to $(\pi,0)$ and $(0,\pi)$ and it reaches
${\bf{k}}=(\pi,3\pi/4)$ and $(3\pi/4,\pi)$ 
for $\langle n\rangle=0.76$ (0.67). The figure clearly shows
that the intensity of the peak in S({\bf{k}}) is drastically reduced as the
doping increases. In fact, when incommensurability appears for positive
$t'$ (Fig.2-h) the peak can barely be seen. However, one has to remember
that these features could be strongly enhanced as the temperature decreases.

We conclude that negative values of
$t'$ seem to favor incommensurability more than positive values do.
For both signs of $t'$ we found that
doping away
from half-filling, S({\bf{k}}) peaks at ${\bf{k}}=(\pi,k_m \pi)$ and $(k_m
\pi,\pi)$ where $0 \leq k_m \leq 1$, 
i.e. the peak does not move along the diagonal contrary to the results
of Ref.\cite {Good,Imada}. 
In Fig.3 we plot $k_m$ vs $\langle n\rangle$ for $t'=-0.2$, 0 and 0.2 and,
for comparison, we also show the experimental results for 
LSCO\cite{aeppli}. It is clear that $t'>0 (<0)$ slightly
decreases (enhances)
incommensurability in comparison with the $t'=0$ case. 
The results for $t'=-0.2$ are closer to the experimental
data but to reach a better agreement, it appears that larger
negative values of $t'$ should be used ($t' \approx -0.4$ linearly
extrapolating the results at $t'=0.2$, 0 and -0.2). This is an example where
the value of $t'$ suggested by comparing Fermi surface shapes
obtained from band structure calculations with the ones obtained in the
non-interacting case ($t'=-0.2$ for LSCO) 
does not provide a good fit for magnetic properties.

We have tried to study the case $t'=-0.45$, which is the value suggested for
YBCO. Due to sign problems, which become worse as $|t'|$ increases, we
have been able to study only the case $U/t=4$ and $T=t/4$ on an $8 \times 8$
lattice. We found that the onset of incommensurability occurs at lower
doping ($\langle n\rangle\approx 0.8$) but the intensity of the peak in the structure
factor, S({\bf{k}}), decreases and it is very difficult to decide
whether incommensurability has developed. This resembles the experimental
results for YBCO\cite{Tran} that do not allow to determine if the peak
in S({\bf{k}}) shifts away from $(\pi,\pi)$. The authors of
Ref.\cite{Si} found a similar behavior.\cite{Sip}

Previous analytical
work\cite{Trem2,Si} predicted that $k_m$ would become different from 1
immediately upon doping for $t'=0$ but only when the density reaches a
critical value
$\langle n_c \rangle < 1$ 
for $t'<0$. Since we work on an $8 \times 8$ lattice we only can detect
incommensurability when $k_m$ jumps from 1 to 0.75 and thus we cannot
check the above mentioned behavior. However, we have not observed 
qualitative differences for the studied values of $t'$. Our
results do not support the claims made in Ref.\cite{Good} where it was
reported that a positive value of $t'$ does not induce incommensurability.
We found that, in this case, a larger hole doping than was studied in
Ref.\cite{Good} is required to observe the effect but the tendency to 
incommensurability eventually occurs.
One of the reasons for the disagreement maybe 
the different symmetry of the clusters studied in Ref.\cite{Good} that may
induce incommensurability along the diagonals rather than towards X and
Y as in the square clusters that we studied,  and as it has been 
experimentally observed. 

To summarize, in this section we 
have studied the behavior of the structure factor as
a function of doping for positive and negative values of the 
diagonal hopping $t'$. We found that $t'$ reduces the strength of the
antiferromagnetic correlations at half-filling. 
Away from $\langle n\rangle=1$ a negative hopping promotes 
incommensuration at a lower doping than a positive one. Using $t'=-0.2$
we did not find good agreement with experimental data for
LSCO showing that deducing bare values of $t'$ based on PES data may be misleading. For $t'=-0.45$ incommensurability was
difficult to observe due to the strong reduction in the intensity of the
peak in the structure factor. This behavior resembles experimental
results for YBCO.

\section{Density versus chemical potential}

A very important property of the Hubbard model is the antiferromagnetic
gap that appears at half-filling. This gap can be observed in the density of
states by studying the spectral function $A({\bf{k}},\omega)$ (see
Sec.VII) or  by analyzing the behavior of the density $\langle n\rangle$ vs $\mu$, 
where $\mu$ is the
chemical potential. It has been found that to change the doping from
holes to electrons, $\mu$ has to cross the gap and a plateau at
$\langle n\rangle=1$ appears in the $\langle n\rangle$ vs $\mu$ curve.\cite{Hub,num} 
Since the gap increases with $U$, so does the size of the plateau.
In this Section we will study the influence of $t'$
on these results. 

Before describing the numerical
data let us discuss the expected behavior in the SDW mean field
approximation. We found that for
the range of parameters analyzed in this paper, 
the value of $\Delta$ that satisfies the mean field equations
(see Appendix)  changes very little when $t'$ is introduced. 
For example, working at $U/t$=4, $\Delta$
is 1.37 for $|t'|=0.2$, i.e. only slightly smaller than at 
$t'=0$ where it takes the
value 1.38. However, the main effect of $t'$ is in a distortion of the bands 
in such a way that the actual gap in the density of states (DOS) 
is smaller than
$\Delta$. For the values of $t'$ studied, this effect is important only
when $\Delta$ is small. 

In Fig.4 we show the mean field energy bands
along $\Gamma-M-X-\Gamma$ in momentum space, 
for $t'=$0 and $-0.2$ and different values of $\Delta$\cite{foot}. 
In Fig.4-a the bands for $U/t=6$ are shown; the gap is equal to $\Delta$
which is
2.48 and it is defined by the difference of energy between the two bands
at ${\bf{k}}=\Sigma=(\pi/2,\pi/2)$ and ${\bf{k}}=X=(\pi,0)$ (actually at
$t'=0$, the energies $E({\bf{k}})$ are degenerate along the line from 
$Y=(0,\pi)$ to $X$). When $t'=-0.2$
the effective gap has been reduced to 2.08 although $\Delta$ remains at
2.48, as can be seen in Fig.4-b, and
it is then $83\%$ of the $t'=0$ value. Now the degeneracy
along $X-Y$ has been removed and the effective gap is defined by $E(X)$ in
the conduction band and $E(\Sigma)$ in the valence band. 
Results for $U/t=4$ are presented in Fig.4-c and d. In this case the effective
gap is reduced from 1.38 to 0.97, being now $70\%$ of the original, when
$t'$ changes from 0 to -0.2. It
is clear that as $U/t$ decreases 
the effect of $t'$ on the size of the gap becomes more important.
In Fig.4-e we present results for $\Delta=0.5$. This value is
interesting because it is well known\cite{Hub} that the effective
size of the gap is reduced in the weak-coupling case due to quantum
fluctuations and this value of $\Delta$ could provide a more accurate
representation of the $U/t=4$ numerical data.\cite{Hub} Note that in this case,
when $t'=-0.2$ the effective gap is very small.
Then we would expect to see the plateau in $\langle n\rangle$ vs $\mu$ very much
reduced. For $U/t=6$, on the other hand, 
the gap changes only slightly at finite $|t'|=0.2$ 
with respect to its value for $t'=0$  
(see Fig.4-a and b). Then, as $U/t$ increases, keeping  $t'$ fixed, 
our expectation is that the effect
of $t'$ on the size of the gap will become increasingly irrelevant.

Now let us analyze the same problem but using QMC
techniques.
In Fig.5-a,b,c we present $\langle n\rangle$ vs $\mu$ for $t'=0$ and $\pm 0.2$ on 
an $8 \times 8$ lattice at $\beta t=6$ in the non-interacting case. 
When $t'$ is negative (positive), half-filling is
achieved for a negative (positive) value of the
chemical potential rather than at
$\mu=0$. It is clear from the figure that, as
we pointed out in Section II, the curve for $t'=0.2$ 
can be obtained from
the curve at negative $t'$ using the prescription
$\langle n\rangle(t',\mu)\rightarrow 
2-\langle n\rangle(-t',-\mu)$ even with interactions present. 
Then, below we will only present results for $t'=-0.2$. 

What is the effect of a finite Coulombic repulsion? 
In Fig.5-d and e  we present $\langle n\rangle$ vs $\mu$ for $U/t=4$ and $\beta t=6$ on an
$8 \times 8$ lattice. When $t'=0$ (Fig.5-d), $\langle n\rangle$ has a plateau
centered at 1 and it is
clear that the chemical potential has to move across the gap to
change the doping from electrons to holes. For $t'=-0.2$ (Fig.5-e) the
plateau is much smaller 
and the type of dopants can be changed from holes to electrons 
varying the chemical potential by a small amount. 
This behavior is in agreement with
experimental data for the high $T_c$ cuprates 
obtained by J.~Allen's group\cite{Allen} who observed
that the chemical potential does not move across the gap as the dopants
are changed from electrons to holes (although
these results have not been yet reproduced using other experimental
techniques). One alternative 
explanation of our numerical results may be that the ground
state in this case is paramagnetic and the antiferromagnetic gap does
not exist, but as we showed in Section III (Fig.1-a) long range
antiferromagnetic order has already developed in our finite cluster for
the couplings we used thus this idea is incorrect. 
The SDW mean
field calculation presented in the Appendix allows us to better 
interpret the data
if we assume that
the antiferromagnetic gap for $U/t=4$ is smaller than the mean field
prediction due to quantum fluctuations. If this is the case, the
effective gap almost disappears as it was shown before in Fig.4-f.
As the Coulombic repulsion is increased to $U/t=6$ the
plateau at $\langle n\rangle=1$ reappears 
for $t'=-0.2$ as can be seen in Fig.5-g. In
this case the plateau is similar to the one observed for $t'=0$
(Fig.5-f). This behavior is also in agreement with the SDW 
mean field results. The curves presented in
Fig.5 appear to be very smooth and they do not indicate the presence of
phase separation in the system.

Summarizing, in this section we studied 
the behavior of the density of electrons as a
function of the chemical potential for positive and negative values of a
diagonal hopping $t'$. We found that $t'$ reduces the size of the
gap. We also found that in weak coupling ($U/t \leq 4$) and at
$|t'|=0.2$ the chemical potential does not move across the naive gap of
order $U$ when the
dopants change from holes to electrons in agreement with the PES results
of Allen et al.\cite{Allen} At larger $U/t$, the effect of $t'$ becomes
less important as observed in previous sections.

\section{Local magnetic moments and double occupancy}

The square of the magnetic moment per site is related to the probability
of double occupancy
$\delta=\langle n_{{\bf{i}}\uparrow}n_{{\bf{i}}\downarrow}\rangle$ 
through the
expression

$$
\langle(m_{\bf{i}}^z)^2\rangle=\langle n\rangle-2\langle n_{{\bf{i}}\uparrow}n_{{\bf{i}}\downarrow}\rangle.
\eqno(3)
$$

In the non-interacting case the probability of double occupancy is
independent of $t'$, and it is given by 
$\langle n_{{\bf{i}}\uparrow}n_{{\bf{i}}\downarrow}\rangle=\langle n\rangle^2/4$. The effect of the
interaction $U$ is to suppress the probability of double
occupancy.\cite{Hub} In the interacting case
$\langle n_{{\bf{i}}\uparrow}n_{{\bf{i}}\downarrow}\rangle$ 
depends on $t'$. As it can be seen in
Fig.6, a negative $t'$ decreases $\delta$ when compared with $t'=0$
while a positive value increases it. At half-filling and 
using the symmetry described in Section II, it can be shown
that the probability of
double occupancy is independent of the sign of $t'$. 
According to Eq.(3) the mean-square
magnetic moment per site will slightly increase for negative
$t'$ and decrease for positive $t'$.

\section{Photoemission experiments and Fermi Surface}

In the last few years much progress has been made in the development of
angle-resolved photoemission (ARPES) techniques and reliable data for the
quasiparticle dispersion near optimal doping for $Bi_2Sr_2CaCu_2O_8$
(Bi2212) and YBCO have been obtained.\cite{dessau} The study of the
quasiparticle peak as a function of energy and momentum provides
the shape of the Fermi surface for the different materials.
Until recently, some 
experiments seemed to provide evidence for large electron-like
FS\cite{campu} but these results have been lately challenged by Aebi et
al.\cite{aebi} Using a new photoemission technique these authors have
reported the presence of a so-called shadow band probably
due to antiferromagnetic
correlations\cite{steph}, 
as well as the existence of features resembling 
hole pockets\cite{dan} in Bi2212 (see Fig.7). 
ARPES also provides information on the
quasiparticle bandwidth which appears to be 
of the order of a fraction of eV independently of the material. 
Note that this
bandwidth is much smaller than the one predicted by LDA
calculations\cite{values} which casts doubts on the practice of
selecting values of $t'$ without properly considering the strong
correlations. Flat regions about X and 
Y have also been found in the hole-quasiparticle dispersion. Recently, Wells
et al.\cite{wells} reported ARPES measurements on the antiferromagnetic
insulator $Sr_2CuOCl_2$. This material is very difficult to dope but the
behavior of the quasiparticle at half-filling is important as a
probe for the microscopic theories for holes in antiferromagnets that have been
proposed. The authors of Ref.\cite{wells} tried to fit their
data with a $t-J$ model with
$J=0.125$eV and found good agreement for the bandwidth but they could not
fit the dispersion along the X-Y line. In a recent paper Nazarenko et
al.\cite{sasha} suggested that the addition to the
$t-J$ model of a $t'$ hopping, which
would be material dependent, could help to improve the fitting. Using
the Born approximation they obtained agreement with experiments along the
X-Y line introducing a $t'=-0.35t$, with $J/t=0.3$ and $J=0.125$ eV. Along
the $X-\Gamma$ line disagreement with the experimental data remained
although along this line a quasiparticle peak in the experiment is not clearly
observed and the error bars are very large.  Using the SDW mean field
approximation described in the Appendix we calculated the dispersion and
we tried to fit the experimental data. Using $U/t$=10 and $t'=-0.2t$ we
found reasonable 
agreement with the experimental result as can be seen in Fig.8.
These results confirm the importance of $t'$ terms to describe the cuprates. 

As it was stated in the Introduction, one 
of the reasons often invoked to introduce a $t'$ hopping in
models for high $T_c$ superconductors is to match the shape of the Fermi
surface to experimental results or to LDA calculations.\cite{troup} 
The non-interacting Hubbard model has a
nested Fermi surface (FS) at half-filling and it becomes large and 
electron-like
immediately upon doping. The addition of a negative $t'$ removes the
nesting at half-filling and, instead, the FS becomes open and
hole-like. Upon further doping it eventually closes but it retains hole-like
characteristics. It becomes electron-like only when the hole doping has been
further increased.\cite{Trem} However, it is possible that the
introduction of electronic interactions could modify substantially the
non-interacting FS.
For example, in the two dimensional Hubbard model, the uniform
magnetic susceptibility always decreases upon doping in the
non-interacting case, but
it increases when interactions are strong enough\cite{sus}
(in agreement with experimental results). 
Thus, it is very important to study how interactions
affect the shape of the FS. 

Since we work on finite lattices
it is difficult to accurately determine the FS of a model. Ideally, the
FS should be obtained examining the spectral function
$A({\bf{k}},\omega)$. However, as we show in Section VII presently the
spectral functions cannot be calculated with enough accuracy and
alternative ways of determining the FS must be considered.
In some previous papers\cite{Hub,stef} the FS was obtained
by calculating the position of the momenta where $n({\bf{k}})\approx 0.5$. 
This approach works well for weakly interacting systems, but it may cause
problems when strong interactions are considered. One can imagine that for a
system with a ``hole pocket'' centered at $\Sigma=(\pi/2,\pi/2)$, 
$n({\bf{k}})$ along the
$(0,0)-(\pi,\pi)$ line will jump from a value larger
than 0.5 to zero as the pocket is encountered increasing the momentum
away from $(0,0)$, defining a first Fermi
surface, and then, at the second FS, it will increase again
but to some value smaller than 0.5. Then the
$n({\bf{k}})\approx 0.5$ criterion is clearly incomplete in strong
coupling because it may miss important structure.
We can avoid 
this problem by assuming that a FS is likely to exist where the
numerically obtained   
$n({\bf{k}})$ changes very rapidly. In fact, since experiments actually
only 
detect crossovers in $n({\bf{k}})$ it is important to know where they
occur in numerical simulations. Thus, a study of the values
of $n({\bf{k}})$ will shed light on possible changes in the shape of the
FS as shown below.

Our mean field analysis predicts strong modifications
in $n({\bf{k}})$ as $U/t$ increases at a fixed $t'$. In particular, 
at half-filling, fixed $t'/t$ and large $U/t$, 
$n({\bf{k}})$ resembles more the $t'=0$ case with the same
$U/t$ than the $t'\neq 0$, $U/t=0$ case. This can be seen from
Eq.(9) in the Appendix. If the gap between the conduction and the
valence band is such that at half-filling the conduction band is
completely filled and the valence band is empty, then $n({\bf{k}})$ is
given by the same analytical expression as for the $t'=0$ model. 
Indeed, this is what we have observed with QMC. In Fig.9-a we
present $n({\bf{k}})$ along the line $X=(\pi,0)$ to $Y=(0,\pi)$,
at half-filling for $U/t=0$, 4 and 6 and $t'=-0.2$. 
The mean field value along this line
agrees with the exact result for $t'=0$ ($n({\bf{k}})=0.5$) 
for the finite values of $U/t$ studied. 
It appears that as $U/t$ increases with respect to $|t'|/t$, 
$n({\bf{k}})$ becomes more and more similar to the $t'=0$ case, i.e. the
concavity  tends to disappear and the curve becomes flatter as $U/t$
increases which is very different from the non-interacting curve as
can be seen in the figure. This result at half-filling is not 
unexpected since as $U/t$ increases, double occupancy decreases, the
diagonal hopping becomes less relevant 
and the spin part of the model becomes effectively
described by a nearest neighbor only 
Heisenberg Hamiltonian (since $J'=4t'^2/U=0.04J<<J$ 
for $|t'|=0.2$), i.e. the frustrating $J'$ term is negligible. 

Now let us 
study what happens away from half-filling. In Fig.9-b, c and d, we
present $n({\bf{k}})$ for the same parameters as in Fig.9-a but for
$\langle n\rangle=0.9$, 0.8 and 0.7. We also show the corresponding mean field
results (open symbols) for the interacting case. Mean field results are
expected to become less reliable as $\langle n\rangle$ moves further away from 1 and
antiferromagnetic correlations decrease. However, the MC points are
always closer to the mean field values than to the non-interacting ones.

For negative $t'$, the low temperature mean field analysis
predicts
the formation of hole pockets upon doping about the
$\Sigma=(\pi/2,\pi/2)$ point. On the other
hand, in the non-interacting case the hole-like FS is 
centered about $M=(\pi,\pi)$. In
addition, in the mean field approximation at zero temperature there are
crossovers in $n({\bf{k}})$ denoted by the crosses in Fig.10-a
where $T=0$ mean field results on a $100 \times 100$ lattice at
$\langle n\rangle=0.9$, $t'=-0.2$ and $U/t=6$ are presented. The full squares
define the actual FS and the crosses signal the points where
$n({\bf{k}})$ changes the most rapidly. 
The doted line indicates the points in momentum space
where $n({\bf{k}})=0.5$. Notice that Fig.10-a resembles the results 
obtained by Aebi and collaborators,\cite{aebi} (see Fig.7) 
if we identify the points of maximum variation in $n({\bf{k}})$
with the experimental FS. 
As the temperature is increased within the mean field 
approximation the Fermi surface around the pockets disappears\cite{dan} and,
instead, rapid crossovers are observed, i.e., the solid squares in
Fig.10-a are replaced by crosses. For $U/t=6$ the pockets remain
observable up to $T/t\approx 1/8$. As the temperature increases 
to $T/t=1/4$, i.e., the value used
in our Monte Carlo simulations, it can be seen in Fig.10-b that the
pockets about $\Sigma$ have disappeared 
and only the lines of crossovers remain. The
open squares indicate the points where $n({\bf{k}})=0.5$. Then, the
$n({\bf{k}})=0.5$ criterion applied to the MF data 
would incorrectly indicate a closed FS which
does not resemble the actual FS at $T=0$ at all! In Fig.10-c we show the
FS at $T=0$ for the non-interacting system. In this case,
there is a jump in $n({\bf{k}})$ from 1 to 0 
at the FS and the two criteria to determine its position
agree. As the temperature increases to $T=1/4$, Fig.10-d, in the
non-interacting case, both criteria still agree but a spurious
crossover is induced along the direction $\Gamma-X(Y)$ due to
temperature effects. However, this variation of $n({\bf{k}})$ is much
smaller than along the directions where a FS exists at $T=0$. 

With this qualitative discussion in mind 
let us analyze our QMC results obtained on an $8 \times 8$
lattice at $T=t/4$ using $U/t=6$ and $t'=-0.2$. We will consider a
system slightly doped away from half-filling with $\langle n\rangle=0.9$. The open squares
in Fig.10-e denote the points where $n({\bf{k}})\approx 0.5$ while the
crosses denote the regions of maximum variation of $n({\bf{k}})$. These
results are in excellent agreement with the mean field calculation on an
$8 \times 8$ lattice at the same temperature. Also notice
that this figure resembles Fig.10-b where mean field results at the
same temperature on a $100 \times 100$ lattice are presented. 

Now let us point out the differences with the non-interacting
case shown in Fig.10-f which is 
equivalent to Fig.10-e but with U=0. Notice that the
criterion $n({\bf{k}})\approx 0.5$, denoted by the open squares,
provides a ``Fermi surface'' identical to the one for the $U/t=6$ case
(Fig.10-e) .
However, the position of the maximum variation of $n({\bf{k}})$ (crosses) is
remarkably different.\cite{foot2} 

Then, one of the main results of this paper 
is  that for intermediate and large values of the
Coulomb interaction it is $incorrect$ to assume that the shape of the
Fermi surface is similar to the non-interacting one. The criterion of
maximum variation of $n({\bf{k}})$ does not support such an assumption. 
In this case, a
better agreement is obtained when comparing numerical data 
with a SDW mean field
calculation. According to this mean field approach, hole pockets about $\Sigma$
should be observed at temperatures as high as $T/t=1/8$. This result is
in agreement with experiments for Bi2212\cite{aebi}.
Unfortunately, for $U/t=6$ in the interesting doping regimes
 ($0.8 \leq \langle n\rangle \leq
0.9$), we are not able to obtain accurate results at temperatures lower
than $T/t=1/4$ because the average value of the sign of the fermionic
determinant becomes very small to directly verify the existence of hole
pockets. 

As a final comment we want to add that a
study of the points where $n({\bf{k}})$ varies more rapidly for the
$t'=0$ Hubbard model provides a result qualitatively similar to the one
displayed on Fig.10-e suggesting that a negative $t'$ term may be dynamically
generated.\cite{dan} 
For positive $t'$ hole pockets appear at $X$, $Y$ and symmetrical points. 

In this section we have measured $n({\bf{k}})$ numerically and we studied its
behavior finding the points in momentum space where it changes more
rapidly. The location of these points for negative values of $t'$
determines a shape very similar to the one experimentally 
obtained in Ref.\cite{aebi}
where the Fermi surface of Bi2212 was studied.

\section{Dynamical properties}

It is very important to understand how the introduction of $t'$ affects
the spectral function $A({\bf{k}},\omega)$. To evaluate this quantity
we used quantum Monte Carlo and the maximum entropy technique.\cite{silver}
In Fig.11 we show $A({\bf{k}},\omega)$ for
$U/t=4$ on an $8 \times 8$ lattice, at $T=t/6$, half-filling, and for
$t'=0$ (Fig.11-a) and -0.2 (Fig.11-b). 
At half filling and $t'=0$ the chemical potential
is located inside the gap. $A({\bf{k}},\omega)$ 
is the same along the
line from $(0,\pi)$ to $(\pi,0)$ in agreement with the SDW mean
field result and with the numerical observation showing that
$n({\bf{k}})=0.5$ along this line. The size of the gap 
agrees with the result displayed in Fig.5-d. When $t'=-0.20$, as
discussed in Section IV, the gap is reduced and, within the accuracy of
the maximum entropy procedure, only a pseudogap is observed (Fig.11-b). This
again, is in agreement with the results presented in Fig.5-e and with
the SDW mean field calculations. Also notice that the chemical potential
does not lie inside the pseudogap. It is clear that the symmetry along
$X-Y$ has been removed. 

The study of the spectral function should, ideally, determine 
the shape of
the Fermi surface. This is done by
observing the momenta where the quasiparticle peak crosses the
chemical potential. The problem that we have observed with this approach
is that the
quasiparticle peak has a finite width which is rather large due to lack
of accuracy of the maximum entropy technique. In fact,
the value of ${\bf{k}}$ at which the peak starts moving across the chemical
potential from the left 
is very different for the value of ${\bf{k}}$ when the peak has finally
moved completely to the right of the 
chemical potential, and this introduces large error bars. In Fig.11-c we
present the spectral function for $U/t=8$, $t'/t=-0.2$ 
at $\langle n\rangle=0.9$ and $T/t=1/2$ on an $8 \times 8$ lattice. Along the
direction $\Gamma-X(Y)$ it can be seen that a broad quasiparticle peak
moves across the chemical potential; this would suggest the existence of
a Fermi surface in this direction as Aebi et al.\cite{aebi} found but
we cannot determine accurately 
at what value of the momentum the actual crossing
occurs. Along the line $X(Y)-M$ there are indications that a Fermi
surface may exist very close to $X(Y)$. These Fermi surfaces would appear
in the regions where we detected the largest variations of $n({\bf{k}})$
(see Fig.10-e). The determination of the FS through the study of the
spectral function was used in Ref.\cite{doug} where the Hubbard model with
$U/t=8$ and $\langle n\rangle=0.87$ was studied at $T/t=1/2$ on 
up to $12 \times 12$
lattices. The conclusion is that, in this case, there is a closed Fermi
surface centered at ${\bf{k}}=(\pi,\pi)$. In Fig.12 we present the shape
of the Fermi surface that we obtained studying the problem on an $8
\times 8$ lattice and using the criteria described in Section VI. The
$n({\bf{k}}) \approx 0.5$ criterion (open squares) gives an
electron-like Fermi surface closed about ${\bf{k}}=(0,0)$ as previously
observed in Ref.\cite{Hub} but we have provided enough evidence that
this criterion is not necessarily reliable.  
The points of
rapid crossover in $n({\bf{k}})$ (stars) provide two surfaces,
similar to the behavior predicted by the SDW mean field (see Appendix
and Section VI) and the experimental data of Ref.\cite{aebi}. 

In the figure we also plot the
results obtained from Fig.1 in Ref.\cite{doug}. The full circles indicate the
values of ${\bf{k}}$ where the maximum of the quasi-particle peak
crosses the chemical potential and the error bars correspond to the
width of the peak. Once the data of Ref.\cite{doug} are supplemented
by the proper error bars, the results are consistent with those obtained
with the maximum variation of $n({\bf{k}})$ criterion.  
However, due to the large error bars it is still not
possible to decide whether the Fermi surface is closed about
$\Gamma$ or $M$. 

In this section we studied how a diagonal 
hopping $t'$ added to the Hubbard Hamiltonian
affects the behavior of the spectral function $A({\bf{k}},\omega)$. We
observed that in weak coupling ($U/t$=4) the gap that appears at
half-filling is reduced in agreement with the results presented in
Section IV. We found that the data are not accurate enough
to allow the unique determination 
of the shape of the Fermi surface. However,
in the strong coupling regime ($U/t=8$) away from
half-filling, we found indications of a Fermi surface consistent with
the one presented in Section VI.

\section{Conclusions}

In this paper we analyzed the 2D Hubbard model with a diagonal hopping
term in the intermediate coupling regime using Monte Carlo techniques. 
We observed that a diagonal hopping $t'$ decreases the strength of
spin-spin correlations. A negative diagonal hopping accelerates the
onset of incommensurability while a positive one retards it. The
intensity of the peak in the structure factor decreases rapidly as the
system is doped away from half-filling but its strength may increase at
lower temperatures. Comparing our numerical results with experimental
data for LSCO, we found that a $t'$ more negative than
the suggested value\cite{values} -0.2 appears to be needed to reach
quantitative agreement.

We found that $t'$ tends to reduce the size of the
antiferromagnetic gap but this effect is important only in the weak
coupling regime. Double occupancy is enhanced (reduced) by a positive
(negative) diagonal hopping. This effect does not occur in the
non-interacting case. 

A study of $n({\bf{k}})$ allows us to define a
new criterion to find probable locations of Fermi surfaces working
numerically 
at finite temperature. Since experimentalists also work at finite temperature
and find the FS as the points in momentum space where $n({\bf{k}})$ has
rapid crossovers, we looked at these points in our numerical data.
We observed that lines of rapid crossovers in $n({\bf{k}})$ 
appear in positions similar to the FS
obtained experimentally at room temperature for Bi2212\cite{aebi}. 
An analogous result 
is obtained for the $t'=0$ case which
indicates that a negative hopping term may be dynamically generated.

A study of the spectral functions show how the addition of a diagonal
hopping term removes the symmetry along the line $X-Y$ and reduces the
size of the gap at half-filling. 

We conclude that it is very dangerous to extract a bare parameter of the
Hamiltonian like $t'$ from experimental (PES) data where renormalizaed
parameters play the important role.

\section{Acknowledgments}

We thank A. Sandvik for providing his maximum entropy code and 
E.~Dagotto, A.~Kampf, Q.~Si, A.M.~Tremblay and L.~Chen
for useful conversations
and suggestions. 
This work is supported by the Office of Naval Research under
grant ONR N00014-93-0495 and ONR N00014-94-1-1031.
We thank SCRI and the Computer Center at FSU for
providing us access to their Cray-YMP supercomputer and also 
ONR for giving us access to their CM5 connection machine.

\section{Appendix: SDW Mean Field} 

The spin density wave (SDW) mean field formalism has been applied to the
Hubbard model\cite{Bob} and also to the $U-t-t'$ model\cite{Hirsch,Chubu}. 
At half-filling the ground state for the $U-t-t'$ model is
antiferromagnetic if $U>U_c$; the critical coupling $U_c$ is a function
of $t'$ and for  $|t'|=0.2$  we found numerically $U_c \approx
2.5$.\cite{agrH} Since we will study U=4 or higher we want to
concentrate in the antiferromagnetic solution. In this case we found two
energy bands given by

$$
E_k^{\pm}=-4 t' cos k_x cos k_y- \mu \pm E_k^0,
\eqno(4)
$$

\noindent where

$$
E^0_k=\sqrt{\epsilon_k^2+\Delta^2},
\eqno(5)
$$

$$
\epsilon_k=-2 t (cos k_x+cos k_y),
\eqno(6)
$$

\noindent and $\Delta$ is obtained by solving the self-consistent
equations:

$$
{1\over{U}}=\sum_{k}{[f(E^-_k)-f(E^+_k)]\over{E^0_k}},
\eqno(7)
$$

\noindent and 

$$
\langle n\rangle= \sum_{k} n(k).
\eqno(8)
$$

\noindent where $f(x)$ is the fermi function given by ${1\over{e^{\beta
x}+1}}$ and

$$
n(k)={1\over{2}}(1-{\epsilon_k\over{E^0_k}})f(E^-_k)+
{1\over{2}}(1+{\epsilon_k\over{E^0_k}})f(E^+_k).
\eqno(9)
$$

\vfil\eject

%
%

{\bf Figure Captions}

\begin{enumerate}

\item
(a) Spin-spin correlation ${\rm C({\bf r})=\langle S^z_{\bf i} S^z_{\bf i+r}\rangle 
(-1)^{|{\bf r}|} }$ for ${\rm U/t=4}$, ${\rm T=t/6}$ on an $8\times 8$
lattice at half-filling for $t'=0$ (open squares) and $t'=-0.2$ (filled
squares). The error bars are of the size of the dots;
(b) Same as (a) for ${\rm U/t=6}$, ${\rm T=t/4}$. 

\item 
(a) Structure factor S({\bf{k}}) 
along the lines $\Gamma-X-M-\Gamma$ for $U/t=6$
and $\beta t=4$ on an $8 \times 8$ lattice at $\langle n\rangle=0.9$ and $t'=-0.2$;
(b) Same as (a) for $t'=0.2$; (c) Same as (a) for $\langle n\rangle=0.8$;
(d) Same as (c) for $t'=0.2$; (e) Same as (a) for $\langle n\rangle=0.7$;
(f) Same as (e) for $t'=0.2$; (g) Same as (a) for $\langle n\rangle=0.6$;
(h) Same as (g) for $t'=0.2$.

\item Value of $k_m$ 
 as a function of doping for $t'=0.2$ (open squares), 0 (triangles) and 
-0.2 (dark squares) for $U/t=6$ on an $8
\times 8$ lattice and $\beta t=4$. 
The crosses are experimental data from Ref.\cite{aeppli}.

\item
(a) SDW mean field energy bands on an $8 \times 8$ lattice along
$\Gamma-M-X-\Gamma$ in momentum space, 
for $\Delta=2.48$ that corresponds to ${ U/t=6}$ for $t'=0$; the 
effective gap is shown with dashed lines;
(b) Same as (a) but for $t'=-0.2$;
(c) Same as (a) but for $\Delta=1.38$ that corresponds to ${ U/t=4}$.
(d) Same as (c) but for $t'=-0.2$;
(e) Same as (a) but for $\Delta=0.5$. 
(f) Same as (e) but for $t'=-0.2$.

\item 
(a) The density $\langle n\rangle$ as a function of $\mu$ on an 
$8\times 8$ lattice for ${ U/t=0}$, ${ T=t/6}$ and $t'=0$. 
(b) Same as (a) for $t'=-0.2$;
(c) Same as (a) for $t'=0.2$; (d) Same as (a) for $U/t=4$;
(e) Same as (b) for $U/t=4$;
(f) Same as (a) for $U/t=6$ with ${ T=t/4}$;
(g) Same as (f) for $t'=-0.2$. The
error bars are smaller than the symbols.

\item Probability of double occupancy $\delta$ as a function of filling
$\langle n\rangle$ on an 8x8 lattice at $\beta=4$ and $t=1$ 
for U=6 and $t'=0.2$ (open squares), 0 (crosses) and -0.2 (filled squares). The
filled triangles are results for U/t=0.

\item Fermi surface for Bi2212 as shown in Ref.\cite{aebi}.

\item Quasiparticle dispersion of the $U-t-t'$ model calculated using a
SDW mean field using $t'=-0.2t$, $U/t=10$ and $U=4$eV (solid line),
$t-J$ results (dotted line) and experimental ARPES results from 
Ref.\cite{wells}. 
Through the relationship $J=4t^2/U$ we found
$J/t=0.4$ and using $t=0.4$eV we get $J=0.16$eV. 

\item  Quantum MC values for 
$n({\bf{k}})$ along $X-Y$ on an $8 \times 8$ lattice at $\beta
t=4$ and $t'=-0.2$ for $U=0$ (crosses), $U=4$ (filled circles) and $U=6$
(filled squares) and mean field results indicated by open circles
($U=4$) and open squares ($U=6$) and connected with dashed lines
for a) $\langle n\rangle=1.0$ (in this case the MF results along the
direction shown are independent of $U$ and open circles only are used), 
b) $\langle n\rangle=0.9$, c) $\langle n\rangle=0.8$, and d) $\langle n\rangle=0.7$.

\item a) Mean field results for $U/t=6$, $T=0$ and $\langle n\rangle=0.9$ on a $100
\times 100$ lattice. The dark squares denote the Fermi surface; the
crosses indicate the values of ${\bf{k}}$ where $n({\bf{k}})$ changes
most rapidly and the dotted line indicates the values of ${\bf{k}}$
where $n({\bf{k}})=0.5$; b) same as a) for $T/t=1/4$. Here, there is no real
Fermi surface and  the values of ${\bf{k}}$
where $n({\bf{k}})=0.5$ are indicated by open squares; c) same as a) for
$U/t=0$;d) same as b) for $U/t=0$; e) Monte Carlo results
for $U/t=6$, $T/t=1/4$ and $\langle n\rangle=0.9$ on a $8
\times 8$ lattice. The
crosses indicate the values of ${\bf{k}}$ where $n({\bf{k}})$ changes
most rapidly and the open squares indicate the values of ${\bf{k}}$
where $n({\bf{k}})=0.5$;f) same as e) for $U/t=0$.

\item a) $A({\bf{k}},\omega)$ for $U/t=4$, $t'=0$, $T/t=1/6$ on an $8
\times 8$ lattice at half-filling. The moments are in units of $\pi/4$.
The dotted line indicates the position of the chemical potential; b)
same as a) for $t'=-0.2$; c) $A({\bf{k}},\omega)$ for $U/t=8$,
$t'=-0.2$, $T/t=1/2$ on an $8
\times 8$ lattice at $\langle n\rangle=0.9$. 

\item Fermi surface for the 2D Hubbard model with $U/t=8$, $\langle n\rangle=0.87$ at
$T/t=1/2$ using the $n({\bf{k}})=0.5$ criterion (open squares and solid
line), the
maximum variation of $n({\bf{k}})$ criterion (stars and dashed line) 
and from the
maximum entropy results of Ref.\cite{doug} (solid circles, with error
bars, and dotted line).

\end{enumerate}


\begin{references}
\bibitem{foot} The results for $t'=0.2$ are obtained by exchanging
valence and conduction bands.
\bibitem{foot2} Notice the resemblance between Fig.10-f and 10-d
indicating that finite size effects are not very strong.
\bibitem{agrH} This result is in agreement with the value reported in
Ref.6.
\end{references}
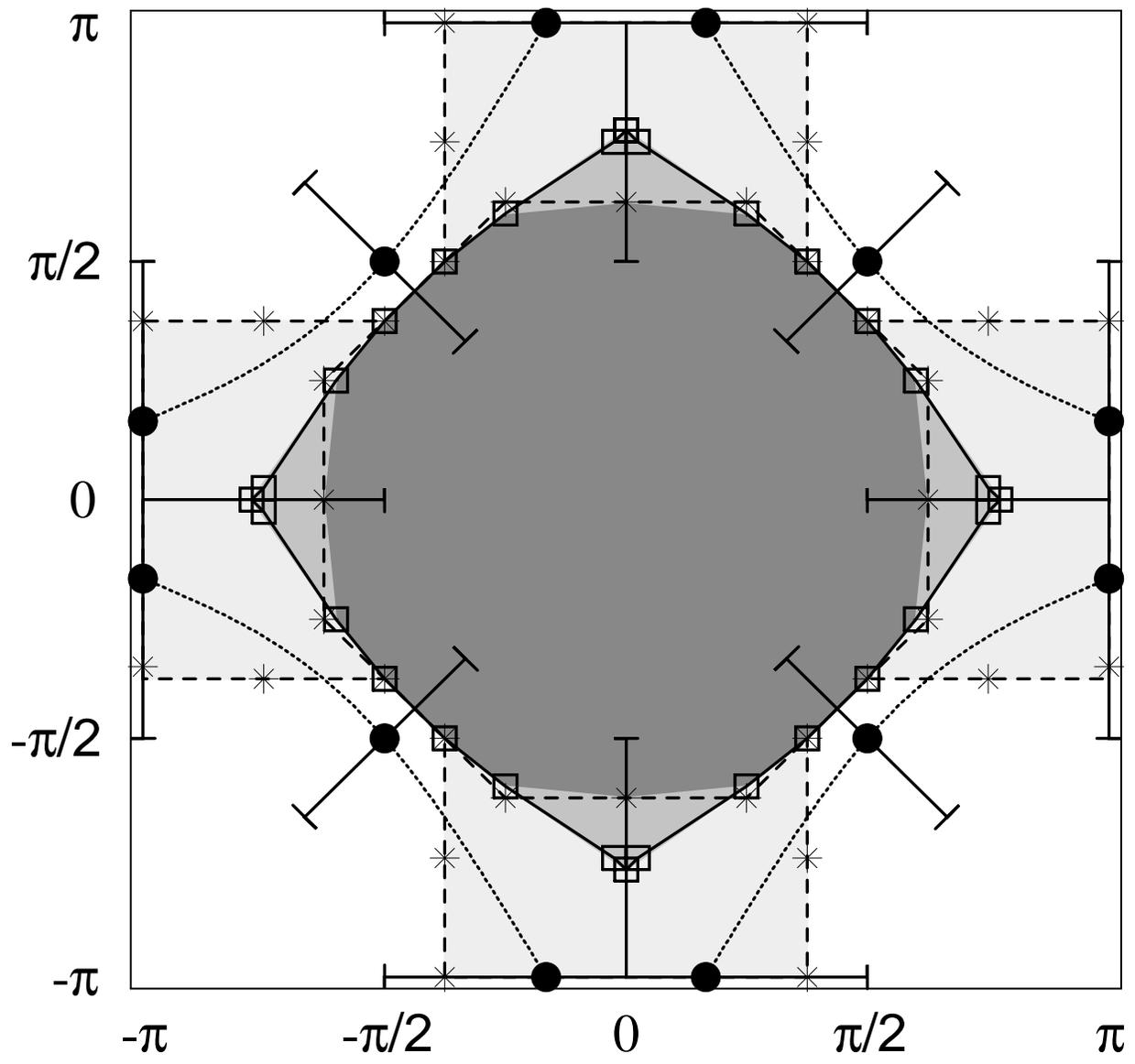
\end{document}